%% file: Main.tex
\documentclass[11pt,draftclsnofoot,onecolumn]{IEEEtran}
\input{Packages}
\IEEEoverridecommandlockouts
\begin{document}
\input{Title}
\begin{abstract}
Four-dimensional Scanning Transmission Electron Microscopy (4D STEM) with data acquired using a defocused electron probe is a promising tool for characterising complex biological specimens and materials through a phase retrieval process known as Electron Ptychography (EP). The efficacy of 4D STEM acquisition and the resulting quality of EP reconstruction depends on the overlap ratio of adjacent illuminated areas. This paper demonstrates how the overlap ratio impacts the data redundancy and the quality of the EP reconstruction. We define two quantities as a function of the overlap ratio that are independent of both the object and the EP algorithm. Subsequently, we evaluate an EP algorithm for varying overlap ratios using simulated 4D STEM datasets. Notably, a 40\% or greater overlap ratio yields stable, high-quality reconstructions.
\end{abstract}
\section{Introduction}\label{sec:introduction}
\begin{figure}[!t]
    \centering\
    \includegraphics[width=0.7\columnwidth]{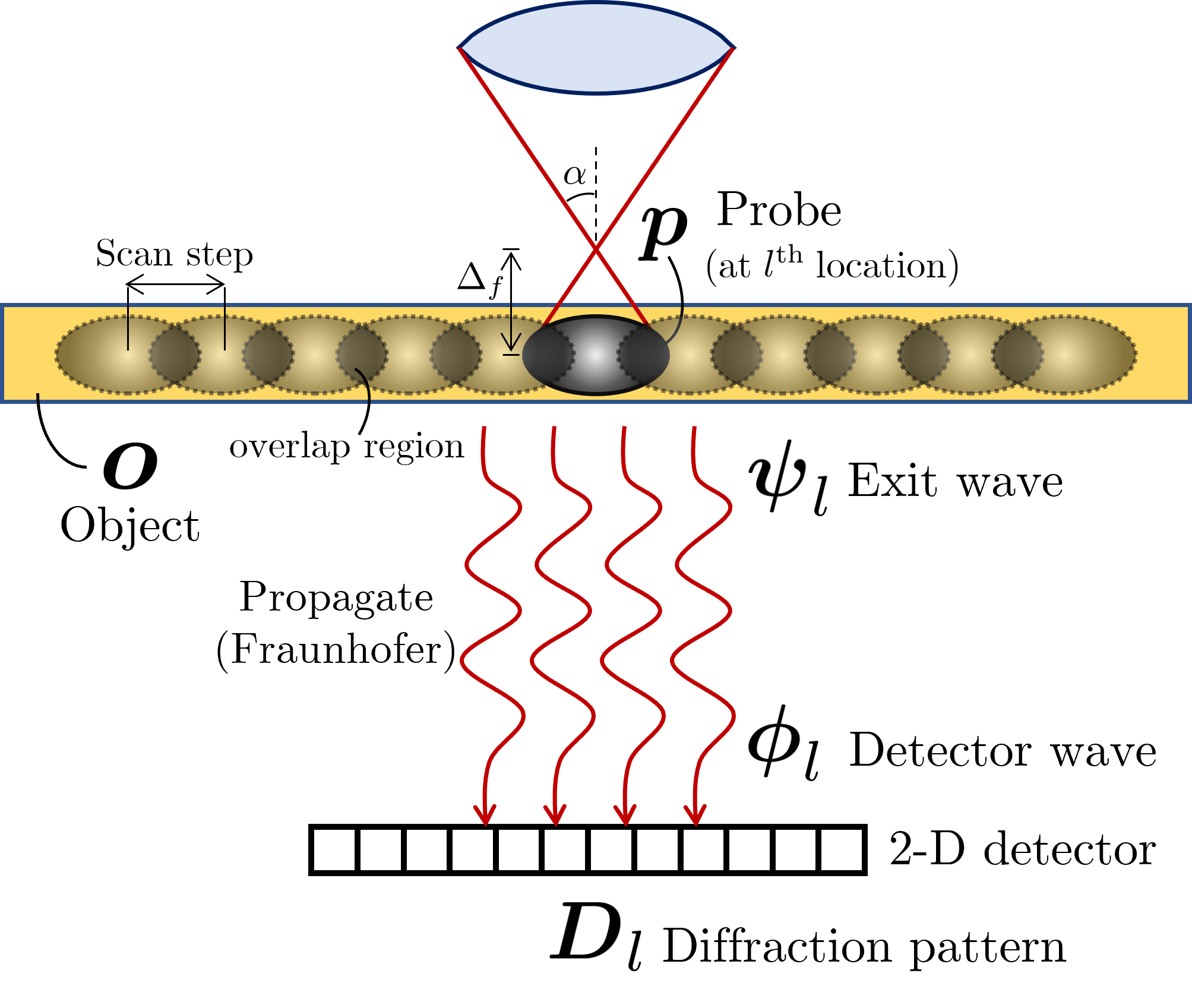}
\caption{\textbf{Defocused-probe 4D STEM.} An electron probe scans a FoV over the object and a diffraction pattern per probe location is collected using a 2-D electron detector. Scanning all probe locations creates the 4-D dataset. Defocus and convergence semi-angle parameters are denoted, respectively, by $\Delta_f$ and $\alpha$ and are discussed in Sec.~\ref{sec:methods}.}
    \label{fig:4d-stem-schematic}
\end{figure}

Four-dimensional Scanning Transmission Electron Microscopy (4D STEM) with a defocused electron probe is a technique that is used to image complex objects structures at atomic resolution~\cite{nellist1995resolution}.
A simplified schematic of defocused-probe 4D STEM is shown in Fig.~\ref{fig:4d-stem-schematic}. An electron probe scans predefined locations across the object, and at each probe position, the object transfer function is encoded in the exit wave. After propagating, \eg generally in the Fraunhofer limit, diffraction patterns are recorded on a two-dimensional (2D) detector, resulting in a 4D STEM data volume.

One of several retrieval algorithms  including \cite{rodenburg2004phase,maiden2009improved,thibault2009probe,luke2004relaxed,pham2019semi,thibault2012maximum,moshtaghpour2023exploring,moshtaghpour2024lorepie} is then applied to the 4D STEM data for the joint recovery of complex-valued object and probe information. The phase of the object, which contains high-resolution details of the object structure, is the primary information recovered. Physical optics show that a diffraction pattern in the far field corresponds to the intensity of the (Discrete Fourier Transform) DFT of the exit wave, meaning that the spatial resolution of the object is determined by the highest angle at which electrons can be collected.

Compared to focused-probe EP~\cite{ophus2019four,robinson2024high,nellist1995resolution}, defocused-probe EP allows imaging of larger Field of Views (FoVs) without compromising the acquisition time by means of increasing the scan step size -- \ie increasing the distance between adjacent probe locations, or equivalently, reducing the overlap ratio between adjacent illuminated regions. 

However, multiple aspects of EP are influenced by the overlap ratio. For instance, the EP problem becomes ill-posed at lower overlap ratios~\cite{song2019atomic}. Conversely, larger overlap ratios allow for imaging larger FoVs but increases  beam-induced damage due to the cumulative radiation dose~\cite{moshtaghpour2024diffusion,jannis2022reducingpart2}. 

In \cite{bunk2008influence}, convergence of PIE with respect to the overlap ratio was studied using simulated and experimental 4D STEM data, and an overlap ratio of 60\% was suggested as a  trade-off between recovery quality and radiation dose. A Fermat spiral trajectory was proposed in \cite{Huang2014optimization} to achieve a uniform and higher overlap ratio for a given number of probe locations. It was experimentally demonstrated in \cite{song2019atomic} that high-quality phase recovery using the extended PIE algorithm depends on both the overlap ratio and acquisition time. The authors in~\cite{bendory2023near,bendory2019blind,jaganathan2016stft,iwen2023phase} have established  with applications in ptychography.

This work further explore the importance of the overlap ratio in EP. Following \cite{bunk2008influence}, we present two approaches to highlight the role of the overlap ratio. First, we analyse the number of times illuminated areas overlap and how often each pixel is illuminated, as a function of the overlap ratio. This approach offers insights that are independent of the phase retrieval algorithm, object structure, electron probe, and detector. Second, we assess the performance of a modified PIE algorithm using simulated 4D STEM datasets. We introduce a framework that ensures consistent dataset creation and reconstruction evaluation for varying overlap ratios and Poisson noise levels.

\section{4-D STEM acquisition model and proposed constrained PIE algorithm}\label{sec:4d-stem}

We consider a STEM scan with data recorded on a 2D pixelated detector with $\Hd \times \Wd$ pixels and an electron probe programmed to scan $\Hp \times \Wp$ positions over the object.

Let $\bs o \in \bb C^{N_{\rm o}}$ be the discretised and vectorised version of the object wave function over a $\Ho \times \Wo$ regular grid, where $N_{\rm o} \coloneqq H_{\rm o}W_{\rm o}$. Similarly, let $\bs p \in \bb C^{N_{\rm d}}$ be the discretised and vectorised version of the probe wave function (see Eq.~\ref{eq:probe-definition}) over $\Hd \times \Wd$ regular grid. Probe locations are indexed by $l \in \dbracket{\Np} \coloneqq \{1,\cdots, \Np\}$, where $\Np \coloneqq \Hp \Wp$ is the total number of probe locations.
For every probe location the exit wave reads

\begin{equation}\label{eq:exit-wave-equation}
    \bs \psi_l = \bs p \odot \bs \Pi_l \bs o \in \bb C^{\Nd}, \quad {\rm for~} l \in \dbracket{\Np},
\end{equation}
where $\odot$ is the Hadamard product and $\bs \Pi_{l} \in \{0,1\}^{\Nd\times \No}$ is a binary matrix that selects only the illuminated region of the object. After propagating in the Fraunhofer limit, the incident wave at the detector, \ie referred to as detector wave,  can be modelled as the DFT of the exit wave ~\cite{hawkes2019springer}:

\begin{equation}\label{eq:detector-wave-equation}
    \bs \phi_l = \bs F \bs \psi_l \in \bb C^{\Nd}, \quad {\rm for~} l \in \dbracket{\Np},
\end{equation}
where $\bs F \in \bb C^{\Nd \times \Nd}$ is the 2D DFT matrix. Finally, the detector records the intensity of the detector wave and we obtain the following simplified sensing model for 4D STEM: 
\begin{equation}\label{eq:4d-stem-equation}
\bs y_l = 
    \left|\bs F (\bs p \odot \bs \Pi_l \bs o) \right|^2 + \bs n_l, \quad {\rm for~} l \in \dbracket{\Np}
\end{equation}
where the operator $\left|\bs \cdot \right|^2$ is applied element-wise and $\bs n_l \in \bb R^{\Nd}$ is a measurement noise. 

Given the 4D STEM data described by \eqref{eq:4d-stem-equation}, the goal of an EP in most practical settings is to recover both the complex-valued object and probe data. However, in this paper, we make two simplifying assumptions: \textit{(i)} the probe data is known from the aberration parameters of the probe-forming system in STEM~\cite{krivanek2017aberration}, and \textit{(ii)} the specimen is a phase object, meaning the amplitude of the object is small and close to one. These assumptions facilitate faster convergence of the EP algorithm.

Following previous work \cite{bunk2008influence}, we base our studies on the PIE algorithm \cite{rodenburg2004phase} and introduce a Constrained PIE (CPIE) algorithm to accommodate the phase object approximation.  As outlined in Algorithm~\ref{alg:pie}, CPIE takes as input a collection of diffraction patterns $\{\bs{y}_l\}_{l=1}^{\Np}$, the probe data $\bs{p}$, the number of iterations $N_{\rm itr}$, the object learning rate $\alpha_{\rm o}$, and an initial object estimate $\bs{o}^{(0)}$, and outputs the recovered object data $\hat{\bs{o}}$. At each iteration, and for a randomly selected probe location, the CPIE algorithm performs the following steps: (line 4) crops the corresponding illuminated region of the object; (line 5) computes an uncorrected exit wave following \eqref{eq:exit-wave-equation}; (line 6) computes an uncorrected wave at the detector plane following \eqref{eq:detector-wave-equation}; (line 7) corrects this wave by replacing the amplitude of the uncorrected detector wave with the square root of the measured diffraction pattern; (line 8) computes the corrected exit wave by inverting \eqref{eq:detector-wave-equation}; (line 9) updates the object estimate; (line 10) enforces the phase object constraint on this new estimate. This final step makes CPIE different from PIE~\cite{rodenburg2004phase} with a negligible computational overhead; (line 11) reinserts the estimated illuminated regions of the object back into the full object FoV.

\setlength{\textfloatsep}{0pt}
\begin{algorithm}[!t]
\caption{Constrained PIE for phase object}\label{alg:pie}
\KwInput{$\{\bs y_l\}_{l=1}^{\Np},  \bs p \in \bb C^{\Nd}, N_{\rm itr}, \alpha_{\rm o}, \bs o^{(0)}$}
\KwOutput{$\hat{\bs o} \leftarrow \bs o^{(N_{\rm itr})}$}
\For(\tcp*[f]{For an iteration}){$t \in \llbracket N_{\rm itr}\rrbracket$}{
$o^{(t)} \leftarrow o^{(t-1)}$\\
\For(\tcp*[f]{For a probe location}){$l \in \llbracket \Np \rrbracket$}{
$\bs o^{(t)}_l \leftarrow  \bs \Pi_l \bs o^{(t)}$ \tcp*[f]{Crop FoV}\\
$\bs \psi^{\rm u}_l \leftarrow \bs p \odot  \bs o^{(t)}_l
$ \!\!\!\tcp*[f]{Uncorrected exit wave}\\
$\bs \phi^{\rm u}_l \leftarrow \bs F \bs \psi^{\rm u}_l$ \tcp*[f]{\!\!\!Uncorrected det. wave}\\
$\bs \phi^{\rm c}_l  \leftarrow \sqrt{\bs y_l}\odot e^{i\angle\,\bs \phi^{\rm u}_l}$  \tcp*[f]{\!\!\!Corrected det. wave}\\
$\bs \psi^{\rm c}_l \leftarrow \bs F^{-1} \bs \phi^{\rm c}_l
$ \tcp*[f]{Corrected exit wave}\\
$\bs o^{(t)}_l \!\leftarrow \bs o^{(t)}_l\! \!+ \!\alpha_{\rm o}\frac{\bs p^*}{\| \bs p\|^2_{\infty}}(\bs \psi^{\rm c}_l - \bs \psi^{\rm u}_l)$\!\!\!\tcp*[f]{\!\!Obj.\! upd.}\\
$\bs o^{(t)}_l \leftarrow e^{i\angle\,\bs o^{(t)}_l}$\tcp*[f]{Phase object constraint}\\
$\bs o^{(t)} \leftarrow \bs \Pi^\top\! \bs o^{(t)}_l + (\bs I_{\No}\! - \bs \Pi^\top\bs \Pi)\bs o^{(t)}$\!\!\tcp*[f]{full \!obj.}
}
}
\end{algorithm}

\setlength{\textfloatsep}{2pt}
We note in Algorithm~\ref{alg:pie}, correcting the wave at the detector plane using the rule in line 7 is suitable for only noiseless settings and will degrade the recovered data in presence of measurement noise.  

\section{Methods: towards an accurate overlap model}\label{sec:methods}
The ultimate goal of this work is to address the question: \textit{What are the necessary and sufficient overlap ratios for successful EP reconstruction?} To explore this, we ignore concerns about beam damage and investigate two simpler questions: \textit{(i)} \textit{How does varying the overlap affect the amount of redundancy in 4D STEM data?} \textit{(ii)} \textit{Can we identify a best-case scenario for EP and its stability with respect to overlap ratio?}
\subsection{A Formal definition of the overlap ratio}\label{sec:overlap-ratio}
In defocused-probe EP, the probe radius is defined as the radius of the illuminated circular region at the specimen. Let $r$ and $d$ represent the probe radius and scan step size, respectively. The overlap ratio, $\rho$, of two adjacent illuminated regions is hence defined as the ratio of the area of their intersection to the area of one of the circular regions. This ratio is expressed as follows:
 \begin{equation}\label{eq:overlap-ratio}
     \rho = R(\gamma) \coloneqq 2 \pi^{-1}(\cos^{-1}(\gamma) - \gamma \sqrt{1-\gamma^2}),
 \end{equation}
 where $\gamma = \frac{d}{2r}$ is the ratio between the scan step size to probe diameter. 

To study the role of the overlap ratio in EP, it is necessary to formulate the inverse of the function $R$. Specifically, given an overlap ratio, we need to compute the parameter $\gamma$. The linear relationship $\rho = 1- \gamma$ proposed  in \cite{bunk2008influence} provides a simple approximation, which we extend. By employing Taylor series expansion techniques, we propose an approximate inverse of the function $R(\gamma)$ in \eqref{eq:overlap-ratio} as
\begin{equation}\label{eq:overlap-ratio-approximation}
    R^{-1}(\rho)\! \coloneqq\!
\begin{cases}
\frac{\pi}{4}(1-\rho), &\!\!\! \rho_1 < \rho \le 1,\\
 \cos\big( \frac{1}{2}(\frac{\pi}{2}\! -\! 1\! +\! \sqrt{2\pi\rho + 3 - \pi})\big),&\!\!\!  \rho_0 \le \rho \le \rho_1,\\
\cos(\frac{1}{2}(6\pi\rho)^{1/3}), &  0\le \rho < \rho_0.
\end{cases}
\end{equation}
Here, $\rho_0$ and $\rho_1$ are threshold parameters that control the accuracy of the proposed approximation. Numerical analysis indicates that setting $\rho_0 = 0.0448$ and $\rho_1 = 0.5816$ results in a sufficiently small error, with $|\rho - R\big(R^{-1}(\rho)\big) | < 0.008$. 

\subsection{Recovery-agnostic analysis}\label{sec:recovery-agnostic}
In this work, the overlap ratio is defined for two adjacent illuminations. Considering one of these as the primary illumination, multiple other illuminations may overlap with this primary region with a given overlap ratio. The first quantity we examine as a measure of redundancy is the number of illuminated regions overlapping with a primary region, denoted by $D(\rho) \in \bb N$. Numerical analysis of $D(\rho)$ is computationally expensive, and we therefore propose a method for fast computation of $D(\rho)$. Let $(l_1,l_2) \in \{0,\cdots,\Hd-1\} \times \{0,\cdots,\Wd-1\}$ be a pair of probe location indices on a 2D grid and consider the middle probe location as the primary illumination. For a fixed $\rho$, let $\gamma = R^{-1}(\rho)$. 
We note that if $\gamma (l_1^2+l_2^2)^{1/2} < 1$, then the $(l_1, l_2)^{\rm th}$ illumination will overlap  the primary illumination. Hence, the quantity $D(\rho)$ can be calculated as
\begin{equation}\label{eq:d-difinition}
    D(\rho) = |\{(l_1,l_2)\ne (0,0) : R^{-1}(\rho)(l_1^2+l_2^2)^{1/2} < 1\}|.
\end{equation}

Furthermore, we define a pixel-wise quantity $C(\rho)$, which counts the number of times each pixel is illuminated during a STEM scan.
From $C(\rho)$ we define $m_C(\rho), M_C(\rho),\mu_C(\rho)$, and $\sigma_C(\rho)$ as the minimum, maximum, mean, and variance of this quantity, respectively. Computing $C(\rho)$ for a full FoV can be computationally expensive. Our approach reduces computational complexity by focusing on the probe diameter $L$. We define two concentric square regions: a simulation box with dimensions $3L \times 3L$ pixels and an evaluation box with dimensions $L \times L$ pixels. For a given overlap ratio, only the former box is simulated and we compute $C(\rho)$ within the evaluation box, which circumscribes the central illumination.

Importantly, the quantities $D$ and $C$ are agnostic to the recovery algorithm, object, and electron probe; and depend only on the geometry of the scan and we now investigate their correlations with the performance of a phase retrieval algorithm. 

\subsection{A best-case EP scenario}\label{sec:recovery-dependant}
In the following, we describe an EP problem that can be considered a best-case practical EP scenario and examine the sensitivity of this problem to overlap ratio.

\sqp\noindent \textbf{4D STEM data generation.}
Probe data is generated according to \cite[Eq.~2.4]{hawkes2019springer}. In its continuous form, the probe function $p(\bs{r})$ in 2D real-space coordinates $\bs{r} = (r_1, r_2)$ is defined as
\begin{equation}\label{eq:probe-definition}
    p(\bs r) =I_0 \cl F^{-1}_{\bs k}\big\{A(\bs k) \exp({i\frac{\pi}{\lambda}\Delta_f |\bs k|^2})\big\},
\end{equation}
where $I_0$ controls the intensity of the probe, $\lambda$ is the wavelength, $\bs{k} = (k_1, k_2)$, with $|\bs{k}| \coloneqq (k_1^2 + k_2^2)^{1/2}$, is the 2D reciprocal-space coordinate, $\Delta_f$ is the defocus as shown in Fig.~\ref{fig:4d-stem-schematic}, and $A(\bs{k})$ is a top-hat aperture function.

A regular probe location grid is created from the overlap ratio and probe radius $r = \tan(\alpha)\Delta_f$. Eq.~\ref{eq:overlap-ratio-approximation} is used to compute $\gamma$ and accordingly the scan step size $d$.

In phase object assumption, we set the amplitude of the simulated object data to a unit value and an arbitrary image is used for the object phase. The object data is extended by symmetric padding, between 75\% to 90\% of the probe array size, ensuring consistent scanning of the central region of the object across different overlap ratios. Finally, diffraction patterns are calculated following the model given in Eq.~\ref{eq:4d-stem-equation}.

\sqp\noindent \textbf{Quality measure.}
CPIE recovers the object data only up to a global phase ambiguity, meaning that the object $\bs o\exp(i\theta)$ for an arbitrary $\theta$ can be a valid solution of the algorithm. For an ambiguity-free evaluation of the recovered object data we use a Normalised Root Mean Square Error (NRMSE) described in \cite{maiden2009improved}, \ie
$$
    {\rm NRMSE}(\bs o, \hat{\bs o}) \coloneqq 10\log\big(\frac{\|\bs o - \nu \hat{\bs o} \|^2}{ \|\bs o\|^2}\big),
$$
where $\nu = \frac{\hat{\bs o}^* \bs o}{\|\hat{\bs o}\|^2}$ allows for multiplication of the object by an ambiguous complex-valued constant, \ie ${\rm NRMSE}(\bs o, c\bs o) = \infty$, for any $c \in \bb C$. Note that only the $\Ho\times\Wo$ central region of the object data, which is extended by symmetric padding, is used for evaluation.

\section{Numerical results}\label{sec:numerical-results}
\begin{figure}[!tb]
    \centering
    \includegraphics[width=0.7\columnwidth]{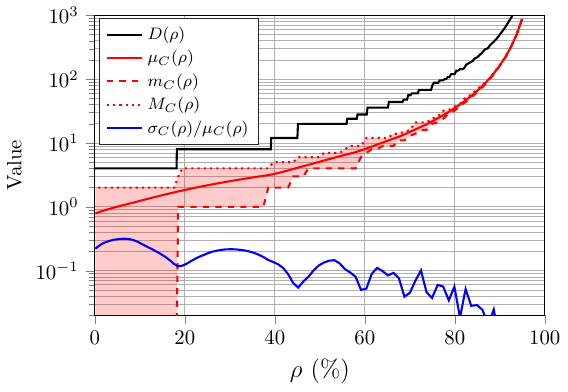}
    \caption{Values of the quantities defined in Sec.~\ref{sec:recovery-agnostic}. These values are independent from the recovery algorithm, illuminated object, and electron probe; and are derived only from the scan geometry, which in this paper follows a raster scan pattern.}
    \label{fig:d_and_c_curve}
\end{figure}
\begin{figure}[tb]
    \centering
    \includegraphics[width=1\columnwidth]{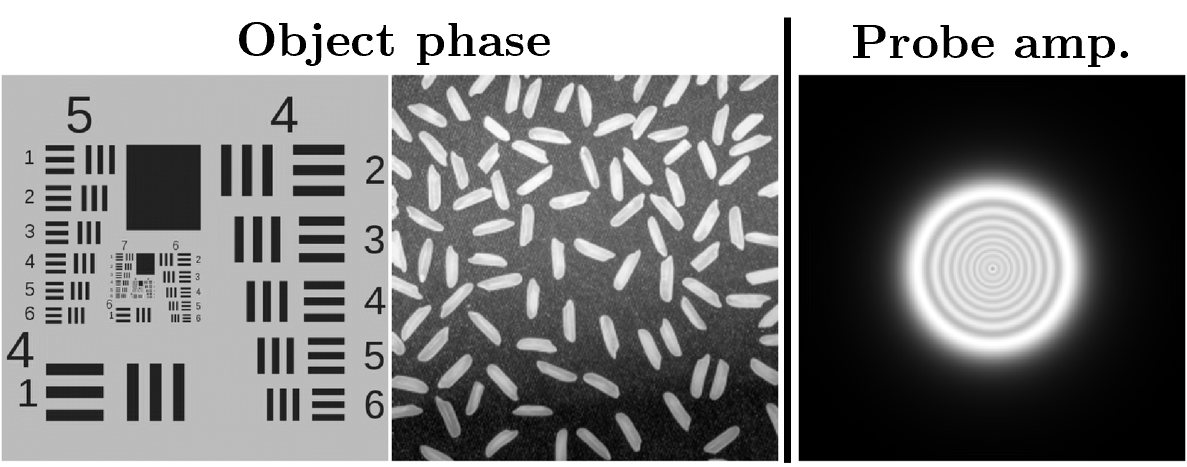}
    \caption{Ground truth images used for generating 4D STEM data. USAF-1951 (left) and rice (middle) images used for the object phase.}
    \label{fig:groundtruth}
\end{figure}
\begin{figure*}[!tbh]
    \begin{minipage}{\linewidth}
    \centering
    \begin{minipage}{0.49\linewidth}
    \centering
    \includegraphics[width=1\columnwidth]{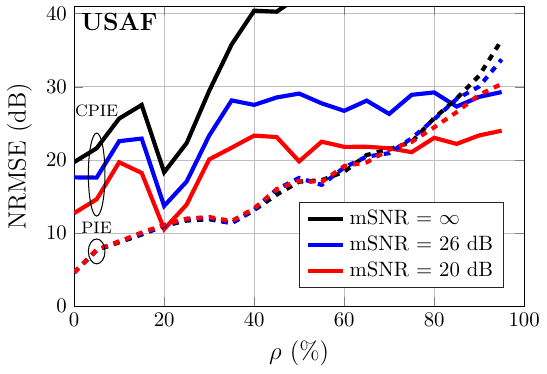}
    \end{minipage}
    \begin{minipage}{0.49\linewidth}
    \centering
    \includegraphics[width=1\columnwidth]{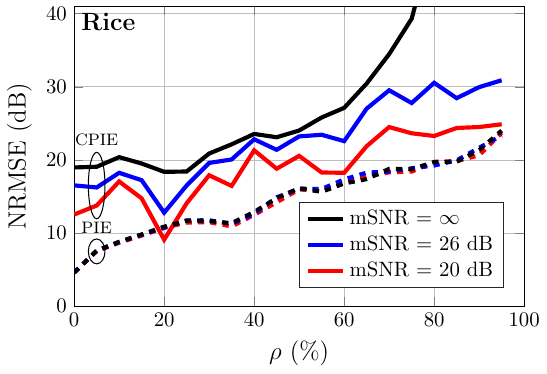}
    \end{minipage}
    \end{minipage}
    \caption{Performance of CPIE and PIE reconstructions for phase object approximation. Efficient overlap ratio for stable reconstruction depends on the EP algorithm and structure of object. For both test images and for considered noise levels, a 40\% overlap yields comparable reconstructions quality to that of 60\% overlap via CPIE algorithm.}
    \label{fig:nrmse_curves}
\end{figure*}
\begin{figure}[!tbh]
    \centering
    \includegraphics[width=0.92\columnwidth]{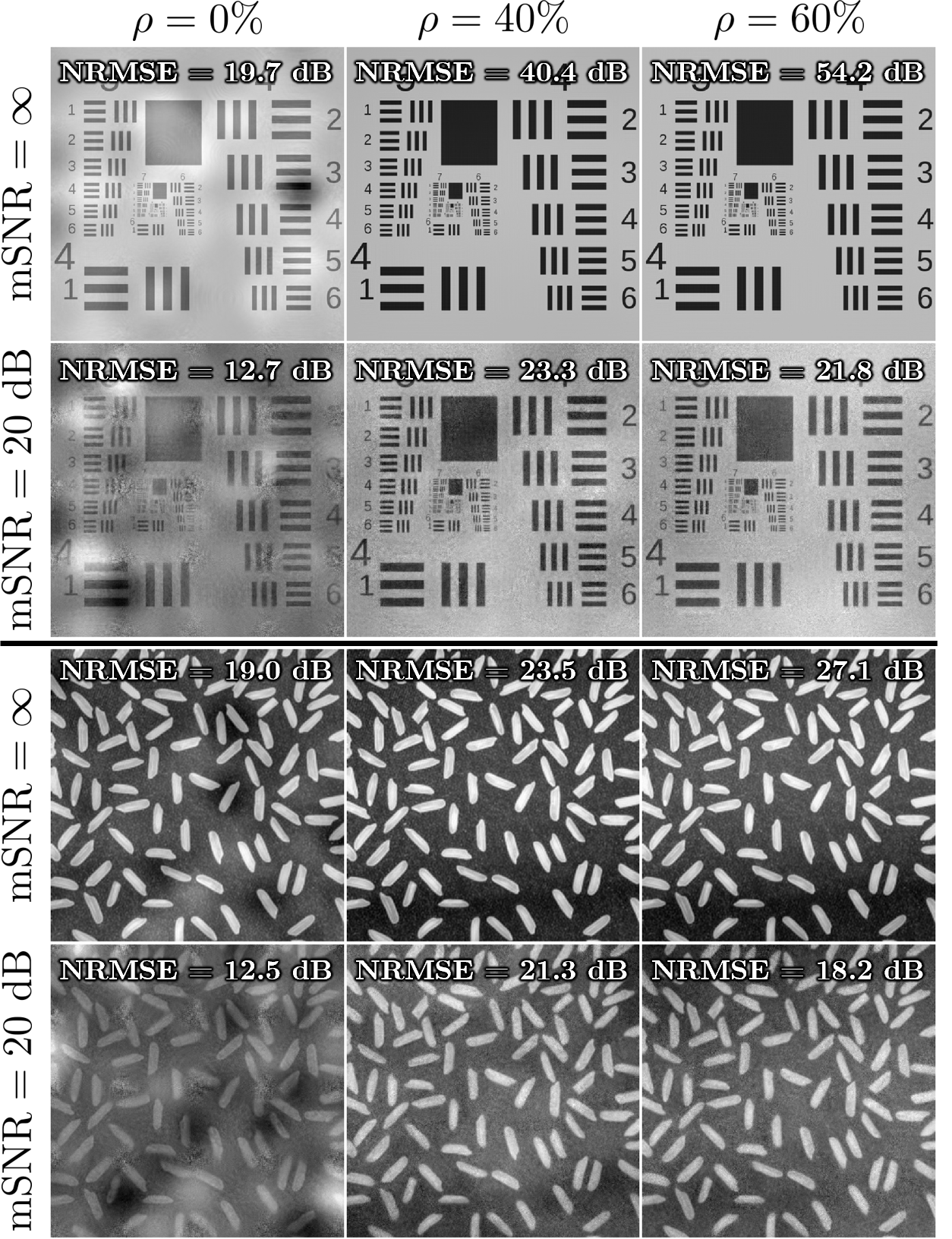}
    \caption{Recovered object phase images via CPIE algorithm for different overlap ratios and mean SNRs.}
    \label{fig:recovered_images}
\end{figure}
Calculated quantities defined in Sec.~\ref{sec:recovery-agnostic} are shown in Fig.~\ref{fig:d_and_c_curve}. We observe that for $18.1\%\le\rho\le 39.1\%$, eight illuminations overlap with the primary illumination and the minimum and maximum number of times a pixel is illuminated is one and four, respectively. Moreover, those values grow exponentially as $\rho$ increases. This would lead to higher-quality phase retrieval but at the cost of a significant increase in beam damage. We hypothesise that a critical overlap ratio for an EP problem may correspond to one of the jump discontinuity points in $D$, $m_C$, or $M_C$ -- for example $\rho \in \{18.2, 39, 45\}\%$ -- as these points indicate significant changes in the amount of redundancy within the 4D STEM data. 

The described best-case EP problem was simulated for the following parameters.
To model the probe, we set $\Delta_f = 1\mu$m, $\alpha = 6$ mrad, and $\lambda = 1.96$ pm. Probe locations were generated with overlaps $\rho \in \{0,5,\cdots, 95\}\%$. For the object phase, two images with $(\Ho, \Wo) = (512, 512)$ pixels were used: \textit{(i)} a rice grain image from MATLAB, and \textit{(ii)} a 1951 USAF resolution test chart. Ground truth probe amplitude and object phase images are shown in Fig.~\ref{fig:groundtruth}. A 2D Detector with $(\Hd,\Wd) = (256, 256)$ pixels was assumed. In addition to simulating a noiseless 4D STEM dataset, we generated two noisy datasets corrupted by Poisson noise (appropriate to an electron detector, especially in electron counting mode~\cite{levin2021direct}) such that the mean Signal-to-Noise Ratio (mSNR) over all diffraction patterns and overlap ratios was ${\rm mSNR} = \{20,26\}$ dB. This was effected by adjusting the intensity of the probe $I_0$. For reconstructions using PIE and CPIE, we set $N_{\rm itr} = 100$, and $\alpha_{\rm o} = 0.1$.

 NRMSE values for two test phase images are shown in Fig.~\ref{fig:nrmse_curves}. We first observe that the PIE algorithm requires large overlap ratio to recover the phase of the object with reasonable quality. For $18.1\%\le\rho\le 39.1\%$, the behaviour of the NRMSE of CPIE correlates with the $\sigma_C(\rho)/\mu_C(\rho)$ curve in Fig.~\ref{fig:d_and_c_curve}. For the USAF and rice images considered here, the CPIE algorithm requires a minimum overlap of 40\% and 70\%, respectively, to achieve phase retrieval quality comparable to that obtained with a 95\% overlap. For both images, an overlap of 40\% yields phase retrieval quality comparable to that of a 60\% overlap. These results also indicate that the rice image presents a more challenging case for the EP problem. This  may arise because the rice image features a non-uniform contrast at low spatial frequency across the FoV.

\section{Conclusions}\label{sec:conclusion}
We derived a model based only on the scan geometry, enabling the identification of critical overlap ratios for ptychographic reconstruction. Our simulations under a phase object approximation showed that a 40\% or greater overlap ratio yields stable, high-quality reconstructions within the parameter space considered.

\bibliographystyle{IEEEtran}
\bibliography{Main}
\end{document}

%% file: Packages.tex
\usepackage{graphicx} 
\usepackage[mode=buildnew]{standalone}
\usepackage{comment}
\graphicspath{{Figures/}}
\usepackage[sort&compress, numbers]{natbib}
\usepackage{amsmath}
\usepackage{array}
\usepackage{url}
\usepackage{amssymb}
\usepackage{stmaryrd}
\usepackage{mathtools}
\usepackage{amsthm}
\usepackage{tabu}
\usepackage{xcolor}
\usepackage{colortbl}
\usepackage{fullpage}
\usepackage[caption=false,font=footnotesize]{subfig}
\usepackage{multirow}
\usepackage{soul}
\usepackage{tikz}
\usetikzlibrary{arrows}

\usepackage{pgfplots}
\pgfplotsset{compat=1.18}
\usepackage{color}
\usepackage{xr-hyper}
\usepackage{hyperref}
\makeatletter
\newcommand*{\addFileDependency}[1]{
  \typeout{(#1)}
  \@addtofilelist{#1}
  \IfFileExists{#1}{}{\typeout{No file #1.}}
}
\makeatother

\usepackage[ruled,linesnumbered]{algorithm2e}

\SetCommentSty{mycommfont}

\SetKwInput{KwInput}{Input}                
\SetKwInput{KwOutput}{Output}              
\SetKwComment{tcp}{}{}

\newcommand{\bs}{\boldsymbol}
\newcommand{\bb}{\mathbb}
\newcommand{\cl}{\mathcal}

\newcommand{\iid}{
    \ifmmode
        \mathrm{iid}%
    \else%
        iid\xspace%
    \fi%
}
\newcommand{\ie}{\emph{i.e.}, }
\newcommand{\eg}{\emph{e.g.}, }

\newcommand{\sqp}{\vspace{2mm}}

\newcommand{\dbracket}[1]{\llbracket #1 \rrbracket}

\newcommand{\Hp}{H_{\rm p}}
\newcommand{\Wp}{W_{\rm p}}
\newcommand{\Np}{N_{\rm p}}

\newcommand{\Hd}{H_{\rm d}}
\newcommand{\Wd}{W_{\rm d}}
\newcommand{\Nd}{N_{\rm d}}
\newcommand{\No}{N_{\rm o}}
\newcommand{\Ho}{H_{\rm o}}
\newcommand{\Wo}{W_{\rm o}}
\renewcommand{\No}{N_{\rm o}}

%% file: Title.tex
\title{On Overlap Ratio in Defocused Electron Ptychography}
\author{\IEEEauthorblockN{
    Amirafshar~Moshtaghpour\IEEEauthorrefmark{1} \thanks{Authors
    acknowledge the support of the RFI, funded by the UKRI, EPSRC. AIK is also funded by the university of Oxford.
    } and Angus~I.~Kirkland\IEEEauthorrefmark{1}\IEEEauthorrefmark{2}
    }
    \\
    \IEEEauthorblockA{
        \IEEEauthorrefmark{1} Rosalind Franklin Institute, Harwell Science \& Innovation Campus, Didcot, OX11 0QS, UK.\\
        \IEEEauthorrefmark{2} Department of Materials, University of Oxford, Oxford, OX2 6NN, UK.
    }
}

\maketitle